\documentclass{bioinfo}
\copyrightyear{2014}
\pubyear{2014}
\usepackage{bm,subfig,placeins,url}

\begin{document}
\firstpage{1}

\title[CheMA 1.0]{Causal network inference using biochemical kinetics}
\author[Oates \textit{et~al}]{C. J. Oates\,$^{1,*}$, F. Dondelinger\,$^{2}$, N. Bayani\,$^{3}$, J. Korkola\,$^{4}$, J. W. Gray\,$^{4}$ and S. Mukherjee\,$^{2,}$\footnote{to whom correspondence should be addressed}}
\address{$^{1}$Department of Statistics, University of Warwick, Coventry, UK.\\
$^{2}$MRC Biostatistics Unit and CRUK Cambridge Institute, University of Cambridge, Cambridge, UK.\\
$^3$Lawrence Berkeley National Laboratory, University of California, Berkeley, USA.\\
$^4$Knight Cancer Institute, Oregon Health and Science University, Portland, USA.}

\history{Received on XXXXX; revised on XXXXX; accepted on XXXXX}

\editor{Associate Editor: XXXXXXX}

\maketitle

\begin{abstract}

\section{Motivation:}
Network models are widely used as structural summaries of biochemical systems.
Statistical estimation of networks is usually based on linear or discrete models.
However, the dynamics of these systems are generally nonlinear, suggesting that suitable nonlinear formulations may offer gains with respect to network inference and associated prediction problems.

\section{Results:}
We present a general framework for both network inference and dynamical prediction that is rooted in nonlinear biochemical kinetics. This is done by considering a dynamical system based on a chemical reaction graph and associated kinetics parameters. Inference regarding both parameters and the reaction graph itself is carried out within a fully Bayesian framework. Prediction of dynamical behavior is achieved by averaging over both parameters and reaction graphs, allowing prediction even when the underlying reactions themselves are unknown or uncertain.  
Results, based on (i) data simulated from a mechanistic model of mitogen-activated protein kinase signaling and (ii) phosphoproteomic data from cancer cell lines, demonstrate that  nonlinear formulations can yield gains in network inference and permit dynamical prediction in the challenging setting where the reaction graph is unknown.

\section{Availability:}
MATLAB R2014a software is available to download from \url{warwick.ac.uk/chrisoates}. 

\section{Contact:} \href{c.oates@warwick.ac.uk}{c.oates@warwick.ac.uk}; \href{sach@mrc-bsu.cam.ac.uk}{sach@mrc-bsu.cam.ac.uk}
\end{abstract}

\section{Introduction}

Statistical network inference techniques are widely used in the analysis of multivariate biochemical data \citep{Sachs, Ellis}.
These techniques aim to make inferences regarding a network $N$ whose vertices are identified with biomolecular components (e.g. genes or proteins) and edges with (direct or indirect) regulatory interplay between those components. 

Algorithms for network inference are typically rooted in linear or discrete models whose statistical and computational advantages facilitate exploration of large spaces of networks  \citep{Werhli}.
On the other hand, when the network topology is known, nonlinear ordinary differential equations (ODEs) are widely used to model biochemical dynamics \citep{Kholodenko,Chen}. 
The intermediate case where biochemical ODE models are employed to select between network models has received less attention.

We propose a general framework called ``Chemical Model Averaging'' (CheMA) that uses biochemical ODE models to carry out both network inference and dynamical prediction. 
In summary, we consider a dynamical system $d \bm{X} / dt = \bm{f}_G(\bm{X},\bm{\theta})$ where the state vector $\bm{X}$ contains the abundances of molecular species, $G$ is a  chemical reaction graph \citep{Craciun}
that characterizes reactions in the system, 
$\bm{f}_G$ is a kinetic model that depends on  $G$ and $\bm{\theta}$ collects together all unknown kinetic parameters. 
A network $N$ is obtained as a coarse summary $N(G)$ of the reaction graph $G$ in which each chemical species appears as a single node and directed edges indicate that the parent is involved in chemical reaction(s) which have the child as product (we make these notions precise below).
Given time-course data $\mathcal{D}$ consisting of noisy measurements of $\bm{X}$, we carry out inference and prediction within a fully Bayesian framework.
Information on the kinetic parameters is integrated through a prior density $p(\bm{\theta}|G)$.
In particular we treat $G$ itself as unknown and make inference concerning it using the posterior distribution,
\begin{eqnarray}
p(G|\mathcal{D}) \propto p(G) \underbrace{ \int p(\mathcal{D}|\bm{\theta},G)p(\bm{\theta}|G) d\bm{\theta} }_{\text{marginal likelihood } p(\mathcal{D}|G)}
\end{eqnarray}
where the marginal likelihood $p(\mathcal{D}|G)$
captures how well the chemical reaction graph $G$ describes data $\mathcal{D}$, taking into account both parameter uncertainty and model complexity. In contrast to linear or discrete models that are motivated by tractability, our likelihood $p(\mathcal{D}|\bm{\theta},G)$ depends on (richer) reaction graphs $G$ and their associated kinetics.

This paper makes three contributions: (1) A general framework for joint network learning and dynamical prediction using ODE models; (2) a specific implementation (CheMA 1.0), rooted in Michaelis-Menten kinetics, that uses Metropolis-within-Gibbs sampling to allow fully Bayesian inference at feasible computational cost; and (3) an empirical investigation, using both simulated and experimental time-course data, of the performance of CheMA 1.0 relative to several existing approaches for  network inference and dynamical prediction.

\begin{figure}[t]
\centering
\includegraphics[width=0.48\textwidth,trim = 5.28cm 16.2cm 5.2cm 4.4cm,clip]{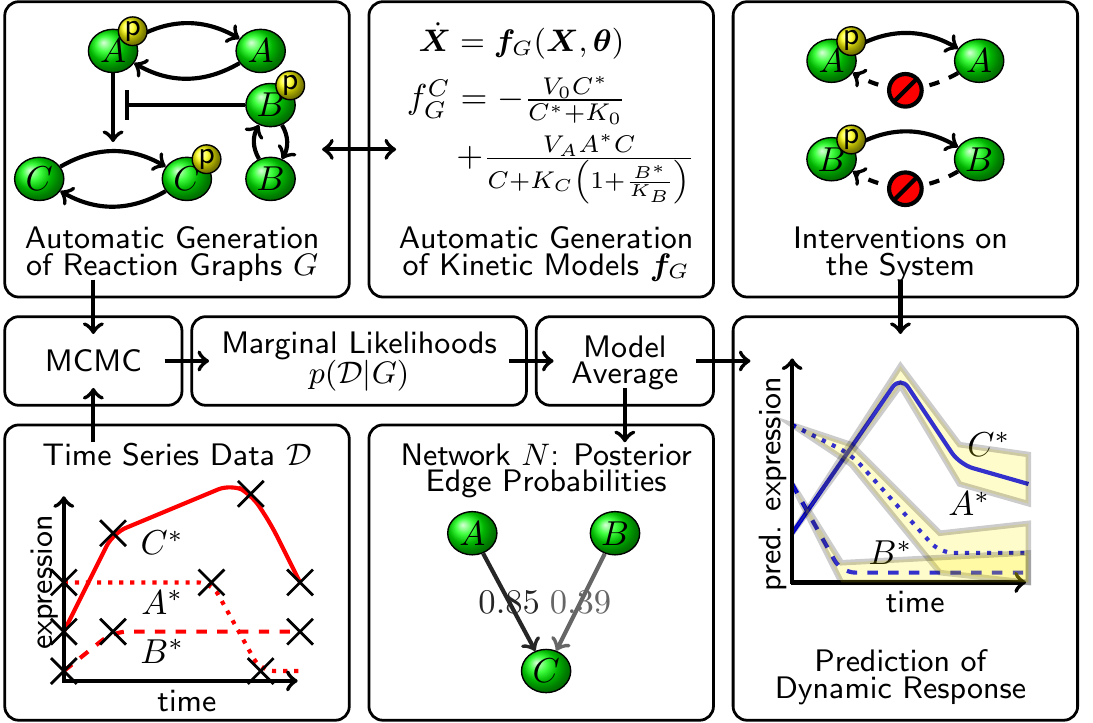}
\caption{Chemical Model Averaging (CheMA). Chemical reaction graphs $G$ summarize interplay that is described quantitatively by kinetic equations $\bm{f}_G$. Candidate  graphs $G$ are scored against observed time course data $\mathcal{D}$ in a fully Bayesian framework.
A network $N$ gives a coarse summary of the system; marginal posterior probabilities of edges in $N$ quantify evidence in favor of causal relationships. 
The reaction graph $G$ (and $N$) is treated as an unknown, latent object and the methodology allows fully Bayesian prediction of dynamics (including under intervention) in the unknown graph setting.}
\label{fig1}
\end{figure}

The statistical connection between {\it linear} ODEs and network inference using linear models has been discussed in \cite{Oates} and exploited in \cite{Gardner,Bansal2}.
Several approaches based on {\it nonlinear} ODEs have been proposed, including \cite{Nachman,Nelander,Aijo,Honkela}. 
Our contribution differs to these by formulating a fully Bayesian approach to both network inference and dynamical prediction that is rooted in chemical kinetics.
Bayesian model selection based on nonlinear ODE has been shown to be a promising strategy for elucidation of specific signaling mechanisms \citep[e.g.][]{Xu}. 
The work we present differs in motivation and approach in that we exploit automatically-generated rather than hand-crafted biochemical models, thereby allowing full network inference without manual specification of candidate models.
\cite{Oates2} performed Bayesian model selection by comparing steady-state data to equilibrium solutions of automatically generated ODE models.
This paper extends this approach to time course data and prediction of dynamics.

There are several considerations that motivate  CheMA:
(i) Inference in biological systems is complicated by strong correlations between components that are co-regulated but not causally linked.
It is well known that, under a  linear formulation, the  causal network $N$ is generally unidentifiable \citep{Pearl}.
For example, it may not be possible to orient certain edges, or edges may be inferred between co-regulated nodes due to strong associations between them.
Nonlinear kinetic equations, in contrast, are able to confer asymmetries between nodes and may be sufficient to enable orientation of all edges \citep{Peters}, although we note that 
nonlinear models still require many assumptions to allow causal inference \citep{Pearl}.
As a consequence, CheMA can in principle aid in causal network inference, and empirical results below support this.
(ii) In contrast to linear models, in CheMA the mechanistic roles of individual variables are respected. This facilitates analysis of data obtained under specific molecular interventions and  enhances scientific interpretability.
(iii) Prediction of dynamical behavior (e.g. response to a stimulus or to a drug treatment) in general depends on the chemical reaction graph.
In  settings  where the graph itself is unknown or uncertain (e.g. due to genetic or epigenetic context), CheMA allows ensemble-averaged prediction of dynamics.

The CheMA framework is general and can in principle be used in many settings where kinetic formulations are available to describe the dynamics,
including gene regulation, metabolism and protein signaling. For definiteness, in this paper we focus on protein signaling networks mediated by phosphorylation and provide a specific implementation of the general framework.
Phosphorylation kinetics have been widely studied \citep{Kholodenko} and  ODE formulations are available, including those  based on Michaelis-Menten kinetics \citep{Leskovac}. 

The remainder of the paper is organized as follows. 
First, we introduce the model and associated statistical formulation. 
Second, we discuss network inference and dynamical prediction within this framework. 
Third, we show empirical results, comparing CheMA 1.0 to several existing approaches, using (i) data simulated from a mechanistic model of mitogen-activated protein kinase (MAPK) signaling, and (ii) phosphoproteomic time course data from human cancer cell lines. 
Finally, we discuss our findings and suggest several directions for further work.

\begin{methods}
\section{Methods}

Below we describe an attempt to implement the CheMA framework, called CheMA 1.0, for the specific context of protein phosphorylation networks.
Fig. \ref{fig1} provides an outline of our workflow below.

\subsection{Automatic Generation of Reaction Graphs $G$}
We construct reaction graphs for $p$ proteins $\{ X_1 \ldots  X_p \} = \mathcal{V}$. Each $X_i$ can be phosphorylated to $X_i^*$; 
the set of phosphorylated proteins is  $\mathcal{V}^*$.
Phosphorylation reactions $X_i \rightarrow X_i^*$ are catalyzed by enzymes $E \in \mathcal{E}_i$; the subscript indicates that each protein  may have a specific set of enzymes (kinases).
We consider the case in which the kinases themselves are phosphorylated proteins, i.e. $\mathcal{E}_i \subseteq \mathcal{V}^*$ (if phosphorylation of $X_i$ is not driven by an enzyme in $\mathcal{V}^*$, we set $\mathcal{E}_i = \emptyset$).
For simplicity we do not consider multiple phosphorylation sites, other post-translational modifications (e.g. ubiquitinylation), protein degradation, nor spatial effects. 
The ability of enzyme $E \in \mathcal{E}_i$ to catalyze phosphorylation of $X_i$ may be inhibited by proteins 
$I \in \mathcal{I}_{i,E} \subseteq \mathcal{V}^*$; the double subscript indicates that inhibition is specific to both target $X_i$ and  enzyme $E$ (see below).

The reaction graph $G$ provides a visual representation of the sets $\mathcal{E}_i$ and $\mathcal{I}_{i,E}$; Fig. \ref{fig1} shows an illustrative example on three proteins A, B and C. A causal biological network $N(G)$ is formed by drawing exactly $p$ vertices and edges $(i,j)$ indicating that $X_i^*$ is either an enzyme catalyzing phosphorylation of $X_j$, or an inhibitor of such an enzyme. 
That is, $(i,j) \in N \iff i \in \mathcal{E}_j \vee \exists E \cdot i \in \mathcal{I}_{j,E}$. 
For the example shown in Fig. \ref{fig1}, the corresponding network $N$ is the directed graph $A \rightarrow C \leftarrow B$.

\subsection{Automatic Generation of Kinetic Models $\bm{f}_G$}
The reaction graph $G$ can be decomposed into local graphs $G_i$ describing enzymes (and their inhibitors) for phosphorylation of protein $X_i$. 
For simplicity of exposition we consider inference concerning $G_i$. Thus, $X_i$ plays the role of the substrate; following conventional notation in enzyme kinetics, we refer to $X_i$ using the symbol $S$.

We use kinetic models $\bm{f}_G$ based on Michaelis-Menten functionals \citep{Leskovac}. 
Here we restrict attention to a relatively simple model class, but more complex dynamics could be 
incorporated if appropriate. 
The rate of phosphorylation due to kinase $E$ is given by $V_E [E] [S]^h/([S]^h+K_E^h)$, which  acknowledges variation of kinase concentration $[E]$ and permits kinase-specific response profiles, parameterized by $K_E$ and $h$, with rate constant $V_E$. In subsequent experiments the Hill coefficient $h$ is taken equal to 1 (non-cooperative binding).
We entertain competitive inhibition, where substrate and inhibitor $I$ compete for the same binding site on the enzyme ($E I \rightleftharpoons  E \rightleftharpoons E S \rightarrow E + S^*$).
When multiple inhibitors are present, they are assumed to act exclusively, competing for the same binding site on the enzyme ($E I \rightleftharpoons  E \rightleftharpoons E I'$), corresponding mathematically to a rescaling of the Michaelis-Menten parameter $K_E \mapsto K_E ( 1 + \sum_{I \in \mathcal{I}_{S,E}} [I] / K_I)$.
We do not model phosphatase specificity; in particular, dephosphorylation is assumed to occur at a rate $V_0 [S^*] / ([S^*]+K_0)$, depending on a Michaelis-Menten parameter $K_0$ and taking a maximal value $V_0$. 

Combining these assumptions produces a kinetic model for phosphorylation of substrate $S$, given by $f_{G,S}(\bm{X},\bm{\theta}_S) =$
\begin{eqnarray}
 - \frac{V_0[S^*]}{[S^*]+K_0} + \sum_{E \in \mathcal{E}_S} \frac{V_E[E][S]}{[S]+K_E \left( 1 + \sum_{I \in \mathcal{I}_{S,E}} \frac{[I]}{K_I} \right)}
\label{MM kinetics}
\end{eqnarray}
where the parameter vector $\bm{\theta}_S$ contains the maximum rates $\bm{V}$ and Michaelis-Menten constants $\bm{K}$ and the (local) graph $G_S$ specifies the sets $\mathcal{E}_S$ and $\mathcal{I}_{S,E}$. 
The complete dynamical system $\bm{f}_G$  is given by taking, for each species $S \in \mathcal{V}$, a model akin to Eqn. \ref{MM kinetics}.
In this way we are  able to automate the generation of candidate parametric ODE models.

\subsection{Model Averaging}
Evidence for a causal influence of protein $i$ on protein $j$ is summarized by 
the marginal posterior probability of a directed edge $(i,j)$ in the network $N$. This is obtained by averaging over all possible reaction graphs $G$, as
\begin{eqnarray}
p((i,j)\in N | \mathcal{D}) = 
\frac{ \sum_{G : i \in G_j}p(\mathcal{D}|G)p(G) }{ \sum_{G} p(\mathcal{D}|G)p(G) }. \label{model average}
\end{eqnarray}
We note that whilst Eqn. \ref{model average} is an intuitive summary statistic, the full posterior over reaction graphs $G$ is also available for more detailed exploration.
In the same vein, model averaging is used to compute posterior predictive distributions (see Supplementary Information).

Following  work in structural inference for graphical models \citep{Ellis} we bound graph in-degree; in particular, we bound the number of kinases $|\mathcal{E}_S | \leq c_1$ and the number of inhibitors $|\mathcal{I}_{S,E} | \leq c_2$  (see Section \ref{sensitivity} below).

Bayesian variable selection requires multiplicity correction to control the false discovery rate and avoid degeneracy \citep{Scott}.
For phosphorylation networks we achieve multiplicity correction using a prior $p(G)$ uniform over the number of kinases, and for a given kinase, uniform over the number of kinase inhibitors:
\begin{eqnarray}
p(G) = \prod_{i=1}^p \binom{p}{|\mathcal{E}_i|}^{-1} \prod_{E \in \mathcal{E}_{i}} \binom{p}{|\mathcal{I}_{i,E}|}^{-1}
\end{eqnarray}
We note that the above prior does not include biological knowledge concerning specific edges; informative structural priors that incorporate biological knowledge are also available in the literature \citep{Mukherjee}.

\begin{figure*}[t]
\centering
\subfloat[]{
\includegraphics[trim = 4.5cm 12cm 3.5cm 4cm,clip,width=0.35\textwidth]{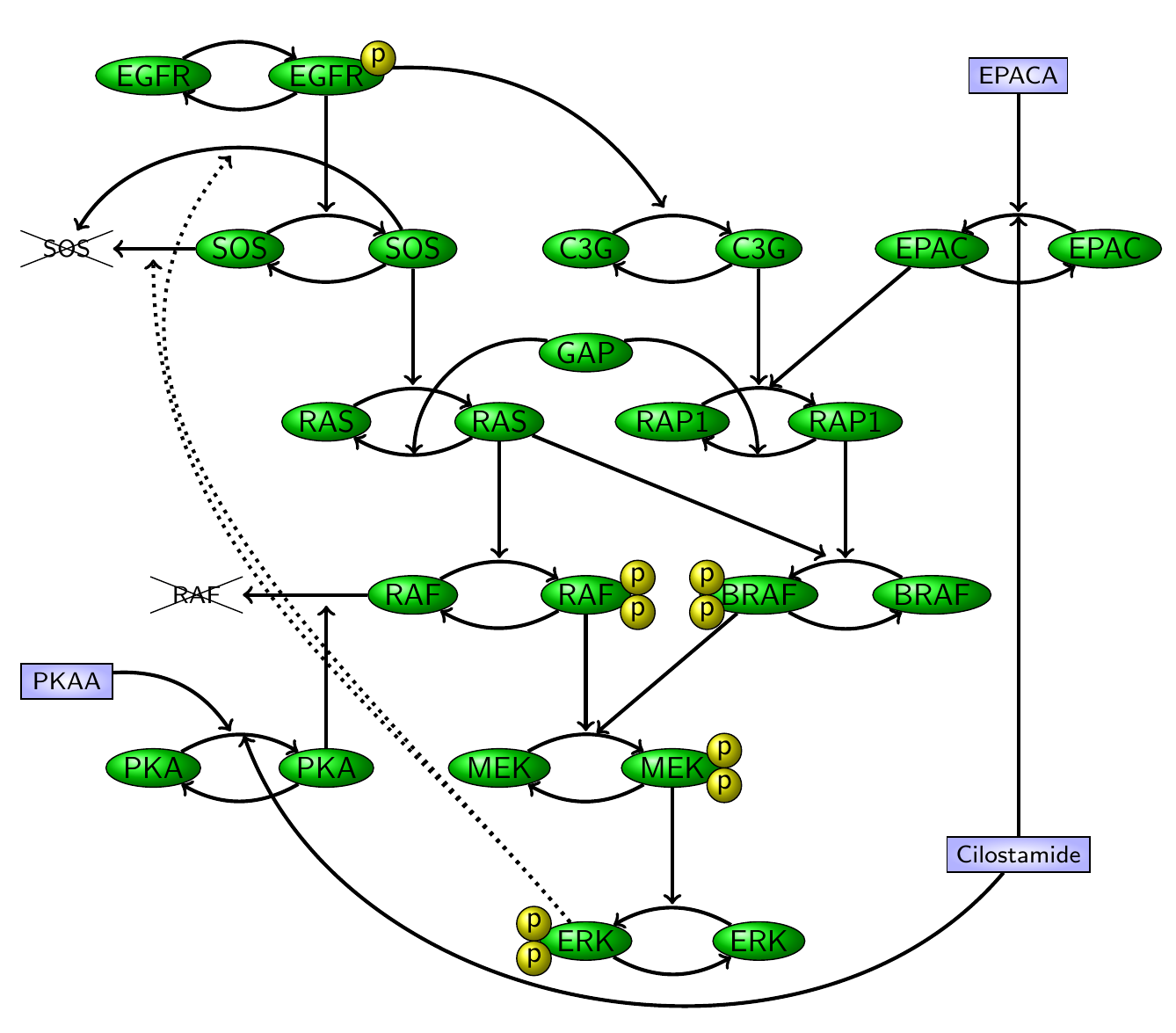} \label{fig2a}}
\subfloat[]{
\includegraphics[width=0.6\textwidth]{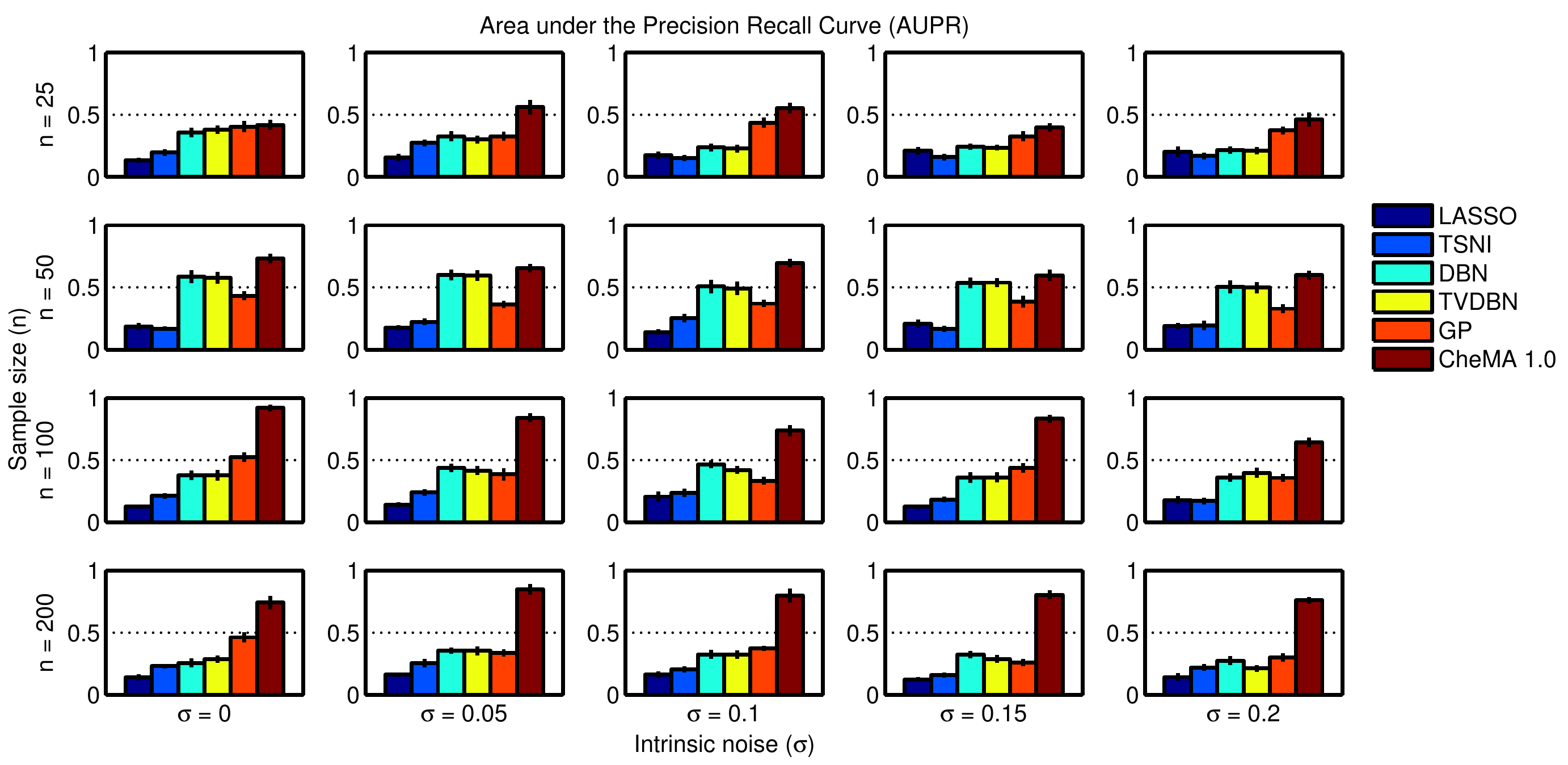} \label{fig2b}}
\caption{Network inference, simulation study. (a) Reaction graph $G$ for the MAPK signaling pathway, due to \cite{Xu}. [The model, based on enzyme kinetics, uses Michaelis-Menten equations to capture a variety of post-translational modifications including phosphorylation.] (b) Area under the precision-recall curve (AUPR; with respect to the true causal network $N(G)$) for varying sample size $n$ and noise level $\sigma$. [Network inference methods: (i) LASSO, $\ell_1$-penalized regression, (ii) TSNI, $\ell_2$-penalized regression, (iii) DBN, dynamic Bayesian networks, (iv) TVDBN, time-varying DBNs, (v) GP, nonparametric regression, (vi) CheMA 1.0, based on chemical kinetic models. 
Error bars display standard error computed over 5 independent datasets. Full details  provided in Supplementary Information.]}
\end{figure*}

\subsection{Statistical Formulation: CheMA 1.0}

The CheMA framework demands a heavily computational approach to inference.
Below we describe an approximate methodology, CheMA 1.0, that aims to accurately approximate the posterior expectations that are of interest.

\subsubsection{Likelihood}
Data $\mathcal{D}$ comprise measurements $y_i(t_j)$ and $y_i^*(t_j)$
proportional to the concentrations of unphosphorylated and phosphorylated forms, respectively, of protein $i$ at discrete times $t_j$, $0 \leq j \leq n$.
Data are scale-normalized to give unit mean for each protein ($\sum_j y_i(t_j) = \sum_j y_i^*(t_j) = n+1$). 
In CheMA 1.0, observables are related to dynamics via ``gradient-matching''. We follow \citet{Bansal2,Aijo,Oates} and employ a simple Euler scheme that approximates the gradient $dX_i/dt$ at time $t_j$ by $z_i(t_j) = (y_i^*(t_j)-y_i^*(t_{j-1})) / (t_j-t_{j-1})$. We note that more accurate approximations could be used, at the cost of requiring more data points or additional modelling assumptions (see Discussion). 
The ODE model $f_{G,S}$ (Eqn. \eqref{MM kinetics}) is formulated as a statistical model by constructing, conditional upon (unknown) Michaelis-Menten parameters $\bm{K}$, a design matrix $\bm{D}_{G,S}(\bm{K})$ with rows
\begin{eqnarray}
\Biggl[ -\frac{y_S^*}{y_S^*+K_0} , \underbrace{\dots ,\frac{y_E^* y_S}{y_S + K_E\left(1+\sum_{I \in \mathcal{I}_{S,E}}\frac{y_I^*}{K_I}\right)}, \dots}_{E \in \mathcal{E}_S} \Biggr]
\end{eqnarray}
and then interpreting Eqn. \eqref{MM kinetics} statistically as 
\begin{eqnarray}
\bm{z}_S & = & \bm{D}_{G,S}(\bm{K}) \bm{V} + \bm{\epsilon}, \; \; \; \bm{\epsilon} \sim \mathcal{N}(\bm{0},\sigma^2\bm{I})
\label{statmodel}
\end{eqnarray}
where $\bm{z}_S = [z_S(t_1), \dots , z_S(t_n)]^T$, $\mathcal{N}$ denotes a normal density, $\sigma^2$ the noise variance,  $\bm{I}$ the identity matrix and, as above, $\bm{V}$ is the vector of maximum reaction rates. 
The appropriateness of normality, additivity and the uncorrelatedness of errors necessarily depends on the data-generating and measurement processes, as well as the time intervals $t_j-t_{j-1}$ between consecutive observations, as discussed in \cite{Oates}.
This approximation has the crucial advantage of rendering the local reaction graphs $G_S$ statistically orthogonal, such that each may be estimated independently \citep[see][]{Hill}.
Iterating over $S \in \mathcal{V}$ permits inference concerning the complete reaction graph $G$.

\subsubsection{Bayesian Inference}
CheMA 1.0 uses truncated normal priors $\mathcal{N}_T(\bm{\mu},\bm{\Sigma})$ with parameters $\bm{\mu},\bm{\Sigma}$ inherited from the corresponding untruncated distribution.
Truncation ensures non-negativity of parameters, whilst normality facilitates partial conjugacy (see below); additional information on truncated normals is provided in the Supplementary Information.
To simplify notation, we consider  a specific variable $S$ and candidate model $G_S$ and omit the subscript in what follows.
In order to elicit hyperparameters $\bm{\mu},\bm{\Sigma}$, we follow \cite{Xu} and assume all processes occur on observable time and concentration scales, that is $\bm{\mu}_{\bm{V}},\bm{\mu}_{\bm{K}} \sim \mathcal{O}(1)$, reflecting that the data are normalized {\it a priori}.
For prior covariance of Michaelis-Menten parameters  $\bm{\Sigma}_{\bm{K}}$ we assume independence of the components $K_i$, so that $p(\bm{K}) = \mathcal{N}_T(\bm{K};\bm{\mu}_{\bm{K}},\nu\bm{I})$, where $\bm{\mu}_{\bm{K}}, \nu$ are hyperparameters.
For the prior covariance $\bm{\Sigma}_{\bm{V}}$ 
of maximum reaction rates 
we take a unit information formulation of the truncated $g$-prior, so that $p(\bm{V}|\bm{K},\sigma) = \mathcal{N}_T (\bm{V};\bm{\mu}_{\bm{V}},n\sigma^2(\bm{D}'\bm{D})^{-1})$ and for the noise parameter we use a Jeffreys prior $p(\sigma) \propto 1/\sigma$.
These latter choices render the formulation partially conjugate, with the conditional density
$p(\bm{V},\sigma|\bm{K},\mathcal{D})$ given in closed form as
\begin{eqnarray}
p(\bm{V},\sigma|\bm{K},\mathcal{D}) & = & \mathcal{N}_T(\bm{V};\bm{\mu},\bm{\Sigma}) \mathcal{IG}(\sigma;a,b),
\label{marg-lik}
\end{eqnarray}
where $\bm{\mu} = \bm{1}/(n+1) + n/(n+1) \times(\bm{D}'\bm{D})^{-1}\bm{D}'\bm{z}$, $\bm{\Sigma} = \sigma^2 n/(n+1) \times (\bm{D}'\bm{D})^{-1}$, $a = (n-1)/2$, $b = (1/2)( \bm{1}'\bm{D}'\bm{D}\bm{1}/n + \bm{z}'\bm{z}-n/(n+1) \times \bm{z}'\bm{D}(\bm{D}'\bm{D})^{-1}\bm{D}'\bm{z})$ and $\mathcal{IG}(\bullet;a,b)$ is an inverse gamma density with shape and scale parameters $a,b$ respectively.

\subsubsection{Marginal Likelihood}
Partial conjugacy of CheMA 1.0 permits an efficient Metropolis-within-Gibbs Markov chain Monte Carlo (MCMC) sampling scheme for the parameter posterior distribution. The conditional density $p(\bm{V},\sigma|\bm{K},G_S,\mathcal{D})$ is given in closed form as in Eqn. \ref{marg-lik} above, while a Metropolis-Hastings acceptance step allows sampling from the remaining conditional $p(\bm{K}|\bm{V},\sigma,G_S,\mathcal{D})$. 
To estimate marginal likelihoods from sampler output we exploit partial conjugacy and use the method of \cite{Chib}, 
evaluating the  identity
\begin{eqnarray}
p(\mathcal{D}|G_S) = \frac{p(\mathcal{D}|\bm{V},\bm{K},\sigma,G_S)p(\bm{V},\bm{K},\sigma|G_S)}{p(\bm{V},\bm{K},\sigma|\mathcal{D},G_S)}
\end{eqnarray}
at particular parameter values $\bm{V},\bm{K},\sigma$ using a Monte Carlo estimate of the posterior ordinate $p(\bm{V},\bm{K},\sigma|\mathcal{D},G_S)$. 
Since inference in CheMA 1.0 decomposes over proteins $X_i \in \mathcal{V}$ and for a given protein, over local models $G_i$, the computations were parallelized (full details and software provided as Supplement).
Alternatively MCMC could be employed over the discrete space of reaction graphs \citep{Ellis} or the joint space of graphs and parameters \citep{Oates2}.

\subsubsection{Interventions on the System}
In interventional experiments, data are obtained under treatments that externally influence network edges or nodes. Inhibitors of protein phosphorylation are now increasingly available; such inhibitors typically bind to the kinase domain of their target, preventing enzymatic activity. We consider such inhibitors in biological experiments below. 
Within CheMA 1.0 we model inhibition  by setting to zero those terms in the design matrix $\bm{D}_{G,S}$ corresponding to the inhibited enzyme $E$ in the treated samples \citep[``perfect certain'' interventions in the terminology of][]{Eaton,Spencer}. 
This removes the causal influence of $E$ for the inhibited samples.

\end{methods}

\section{Results}

\subsection{Hyperparameter Specification and Sensitivity} \label{sensitivity}
For CheMA 1.0 we set hyperparameters $\bm{\mu}_{\bm{V}} = \bm{\mu}_{\bm{K}} = \bm{1}$, $\nu = 1/2$, and the maximum in-degree constraint $c_1 = 2$; we investigated sensitivity by varying these parameters within (a) a toy model of signaling (SFig. 3a-c) and (b) in a subset of the simulations reported below (SFig. 2).
Since the action of inhibition is second-order in the Taylor expansion sense, inference for inhibitor variables $\mathcal{I}_{S,E}$ may be expected to require substantially more data. Indeed, ``weak identifiability'' of second order terms in this context was also reported in \cite{Calderhead}.
A preliminary investigation based on a toy model of signaling revealed that inference for inhibitor sets $\mathcal{I}_{S,E}$ from typical sample sizes was extremely challenging (SFig. 3(d)).
Combined with computational considerations, we decided to fix $c_2 = 0$ for subsequent experiments; that is, we did not include inhibitory regulation in the reaction graph. 
Further diagnostics, including MCMC convergence, are presented in the Supplementary Information.

\begin{figure*}[t]
\centering
\subfloat[]{
\includegraphics[clip,trim = 35cm 10cm 75cm 3cm,width = 0.69\textwidth]{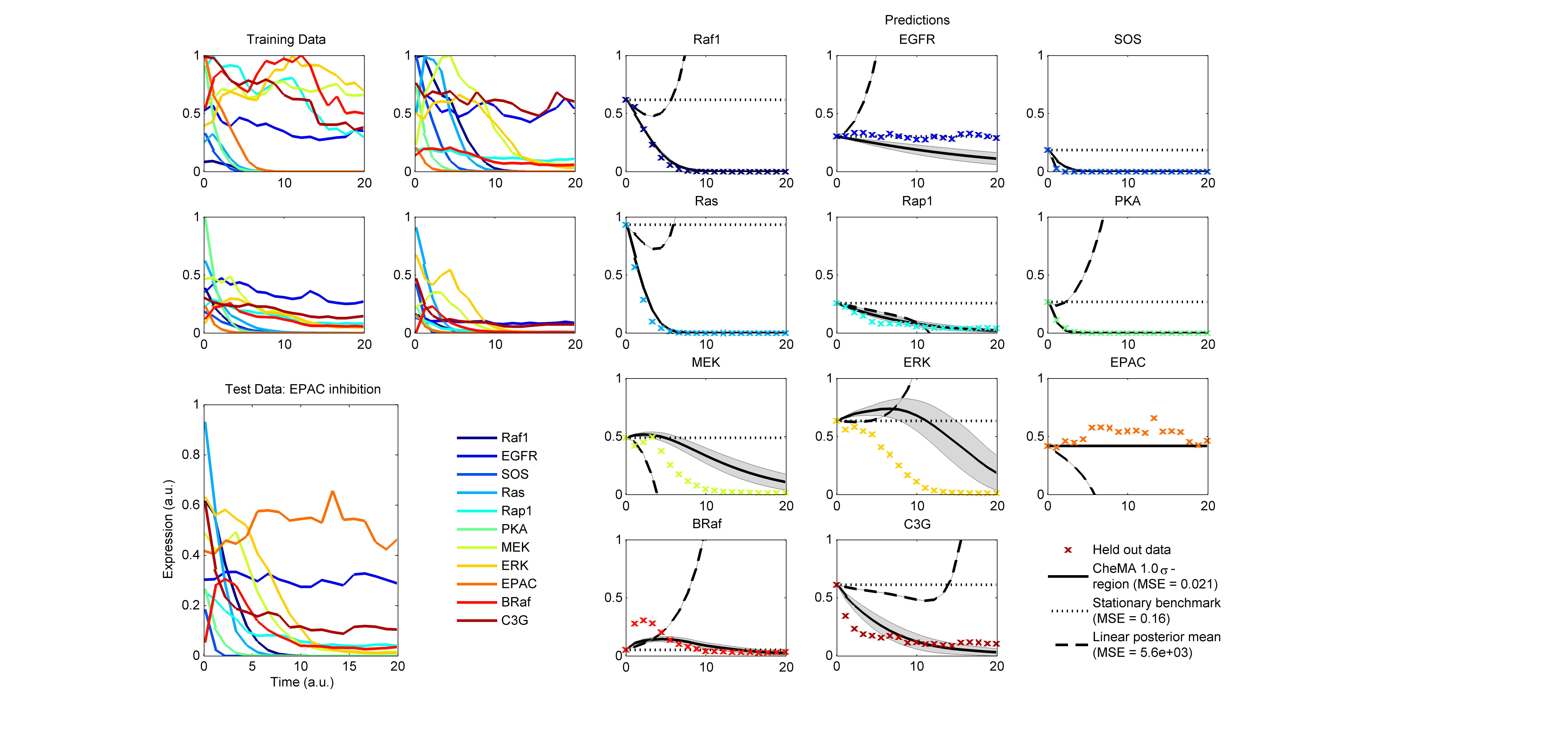} \label{fig3}}
\subfloat[]{
\includegraphics[width = 0.27\textwidth]{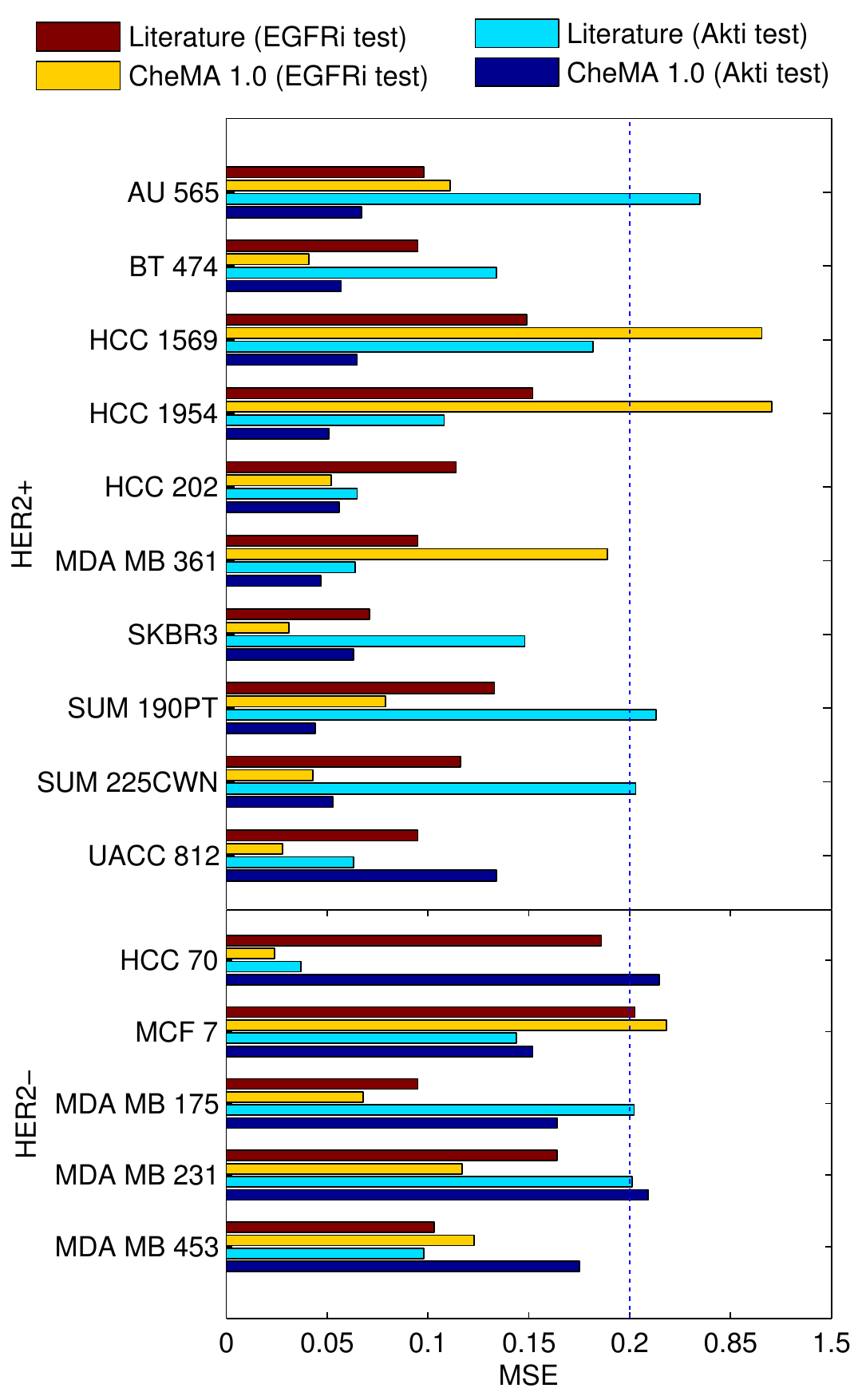} \label{fig4}}
\caption{Predicting dynamical response to a novel intervention: (a) Predicting the effect of EPAC inhibition under the data generating model of \cite{Xu}. 
[CheMA (solid) regions correspond to standard deviation of the posterior predictive distribution.
Linear (dashed) replaces the nonlinear chemical kinetic models with simple linear models.
The stationary benchmark (dotted) simply uses the initial data point as an estimate for all later data points.  
The true test data are displayed as crosses.
Here $n = 100$, $\sigma = 0.1$.]
(b) Assessing prediction over a panel of 15 breast cancer cell lines. 
[Training data were time series under treatment with a single inhibitor; test data represented a second, held-out inhibitor. 
Normalized mean squared error (MSE) was averaged over all protein species and all time points.]}
\end{figure*}

\subsection{In Silico MAPK Pathway}

Data were generated from a mechanistic model of the MAPK signaling pathway due to \cite{Xu}, specified by a system of 25 ODEs of Michaelis-Menten type whose reaction graph is shown in Fig. \ref{fig2a}.
This archetypal protein signaling system provides an ideal test bed, since the causal graph is known and the model has been validated against experimentally obtained data \citep{Xu}. 
Following \cite{Oates} the Xu {\it et al.} model was transformed into an SDE with intrinsic noise $\sigma$. 
Full details of the simulation set-up appear in Supplementary Information.

For inference of the network $N(G)$, we compared our approach to existing network inference methods that are compatible with time course data: 
(i) $\ell_1$-penalized regression (``LASSO") , (ii) 
Time Series Network Identification \citep[``TSNI";][; this is based on $\ell_2$-penalized regression]{Bansal2}, 
(iii) dynamic Bayesian networks \citep[``DBN";][]{Hill}; 
(iv) time-varying DBNs \citep[``TVDBN'';][]{Dondelinger} and
(v) nonparametric (Gaussian process) regression with model averaging \citep[``GP";][]{Aijo}.
Approaches (i-iii) are based on linear difference equations, (iv) relaxes the linear assumption in a piecewise fashion, whereas (v) is a semiparametric variable selection technique. 
We note that since TSNI cannot deal with multiple time courses we adapted it for use in this setting.
Implementation details for all methods may be found in the Supplementary Information.

To systematically assess estimation of network structure we computed the average  area under the precision-recall (AUPR) and receiver operating characteristic (AUROC) curves.
Fig. \ref{fig2b} shows mean AUPR for all approaches, for 20 regimes of sample size $n$ and noise $\sigma$. 
CheMA 1.0 performs consistently well in all regimes, and outperforms (i-v) substantially at the larger sample sizes.
It is interesting to note that the linear and piecewise linear DBNs (iii-iv) perform better at moderate sample sizes compared to higher sample sizes, possibly due to inconsistency arising from model misspecification. 
AUROC results (SFig. 6) showed a broadly similar pattern, with CheMA 1.0 offering gains at larger sample sizes.
For the kinetic parameters, however, we found that CheMA 1.0 struggled to precisely recover the true values $\bm{\theta} = \{\bm{V},\bm{K}\}$, even when the reaction graph $G$ was known (SFig. 8). 
The posterior distribution over rate constants $\bm{V}$ was much more informative than the posterior distribution over Michaelis-Menten parameters $\bm{K}$, consistent with the ``weak identifiability'' of kinase inhibitors that we found in Sec. \ref{sensitivity}.

To investigate dynamical prediction in the setting where neither reaction graph nor  parameters are known, we generated data from an unseen intervention and assessed ability to predict the resulting dynamics (details of the simulation are included in the Supplement).
To fix a length scale, both true and predicted trajectories were normalized by maximum protein expression in the test data.
The quality of a predicted trajectory was then measured by the mean squared error (MSE) relative to the (held out) data points.
The network inference approaches (i-v) above 
cannot be directly applied for prediction in this setting (although they could in principle be adapted to do so).
We therefore compared CheMA 1.0 with the analogous linear formulation, that replaces Eqn. \ref{MM kinetics} by $f_{G,S}(\bm{X},\bm{\theta}_S) = \beta_0 + \sum_{E \in \mathcal{E}_S} \beta_E [X_E^*]$ (see Supplementary Information for details), along with a simple, baseline estimator (the ``stationary benchmark'') that presumes protein concentrations do not change with time.
Fig. \ref{fig3} displays predictions for the dynamics that result from EPAC inhibition. Here CheMA 1.0 provides qualitatively correct prediction, whereas the linear analogue rapidly diverges to infinity (due to poorly estimated eigenvalues).
We therefore focused only on short term prediction, specifically the first 25\% of the time course, for which linear models may yet prove useful.
Over all simulation regimes and experiments, including at small sample sizes, we found that our approach produced significantly lower MSE than both the linear and benchmark models ($\text{MSE}_{\text{CheMA 1.0}} = 0.061$, $\text{MSE}_{\text{Lin.}} = 2.55$, $\text{MSE}_{\text{Bench.}} = 0.199$). Furthermore CheMA 1.0 consistently produced lowest MSE at all fixed values of $n$ and $\sigma$ (SFig. 10; $p < 0.001$ binomial test).

\subsection{{\it In Vitro} Signaling}
Next, we considered experimental data obtained using reverse-phase protein arrays \citep{Hennessy} from 15 human breast cancer cell lines, of which 10 were of HER2+ subtype \citep{Neve}. These data comprised observations for key phosphoproteins Akt, EGFR, MEK, GSK3ab, S6, 4EBP1 and their unphosphoryated counterparts. 
Data were acquired under pretreatment with  inhibitors Lapatinib (``EGFRi"; an EGFR/HER2 inhibitor), GSK690693 (``Akti''; an Akt inhibitor) and without inhibition (DMSO) at 0.5,1,2,4 and 8 hours following serum stimulation, giving a total of $n=15$ observations of each species in each cell line (see Supplementary Information for full experimental protocol).

At present, inferred network topologies for the cell lines cannot be rigorously  assessed since the true cell line-specific networks are not known. 
Inferred topologies partially concord with known signaling (SFig. 11), but the latter is based mainly on studies using wild type cells and may not reflect networks in genetically perturbed cancer lines.
Therefore, for an unbiased test, we considered the problem of prediction of trajectories under an unseen intervention.
We sought to compare performance of CheMA 1.0 against a literature-based ODE model (reaction graph $G$ fixed according to literature and dynamics $\bm{f}_G$ as described above) fitted to training data.
No prior information concerning specific chemical reactions was provided to CheMA 1.0.
This problem is highly nontrivial due to the small sample size, uneven sampling times and the complex observation process associated with proteomic assay data.

Training on DMSO and EGFRi (or AKTi) data, we assessed ability to predict the full dynamic response to Akt (or EGFR) inhibition. In this way, each held-out test set contained trajectories under a completely unseen intervention.
By considering all 15 cell lines, giving 30 held-out datasets, we found that in 19 out of 30 prediction problems CheMA 1.0 outperformed the literature predictor (Fig. \ref{fig4}). 
As for the simulated data, the linear model was not well-behaved for prediction (SFig. 12) and is not shown. 
In the Akti test, of the 10 HER2+ cell lines 9 were better predicted by CheMA 1.0 compared to literature prediction ($p = 0.01$, binomial test; $\text{MSE}_{\text{CheMA 1.0}} = 0.064$ vs $\text{MSE}_{\text{Lit.}} = 0.274$). 
Conversely 4 out of 5 HER2- lines were better predicted by literature ($\text{MSE}_{\text{Lit.}} = 0.145$ vs $\text{MSE}_{\text{CheMA 1.0}} = 0.240$), suggesting that signaling network topology in HER2+ lines may differ to the (wild-type) literature topology.
This agrees with biological understanding of HER2+ cell lines \citep{Neve} and is encouraging from the perspective of CheMA, since {\it a priori} it is far from clear whether the training data, which involved only $p = 6$ species and $n = 10$ data points, contain sufficient information to predict the effect of an unseen intervention, even approximately. However, in two of the failure cases (HCC 1569, HCC 1954; EGFRi test) CheMA 1.0 produced extremely poor predictions ($\text{MSE}_{\text{CheMA 1.0}} > 1$), likely due to the small training sample size.

\section{Discussion}

We proposed a general framework for using chemical kinetics for network inference and dynamical prediction.
The use of chemical kinetics can be expected to contribute gains in causal inference since the underlying models are not structurally symmetric, allowing causal directionality to be established \citep{Peters}.
In empirical results we found that whilst CheMA 1.0 struggled to identify kinetic parameters from data, it was nevertheless able to identify the causal network; this discrepancy is explained by the fact that the latter is in a sense a projection of the former, and can be identifiable even when the full set of parameters are not.

An important challenge in systems biology is to predict the effect on signaling of a novel intervention, such as a drug treatment. 
At present dynamical predictions in systems biology require a known chemical reaction graph, for instance taken from literature; a  system of ODEs is usually specified based on such a graph and used for prediction. However in many settings, the chemical reaction graph may differ depending on (e.g.) cell type or disease state 
and cannot be assumed known. In contrast, CheMA shows how prediction of dynamical behavior  may be possible even when the reaction graph itself is entirely unknown {\it a priori}.
Unlike more convenient linear or discrete formulations \citep[e.g.][]{Maathuis}, our use of chemical kinetic models provides interpretable predictions. For example the dynamic behavior of phosphoprotein concentrations obtained under our methodology are physically plausible (i.e. smooth, bounded and non-negative). 
Furthermore, by averaging predictions over reaction graphs, our approach should provide robustness in (typical) situations where it is unreasonable to expect to identify $G$ precisely.

Several improvements can be made to the CheMA 1.0 
implementation reported here, of which we highlight two:
(i) Gradient matching (rather than numerical solution of the automatically-generated dynamical systems) can help to relieve the computational demands associated with exploration the large model spaces, 
but the Euler approximations we used for this purpose are crude.
Improved gradient matching should be possible (at the expense of requiring more time points) via higher-order expansions, or (at the expense of additional modeling assumptions) kernel regression, the penalized likelihood approaches of \cite{Ramsay,Gonzalez}, or the Bayesian approach of \cite{Dondelinger2}.
(ii) CheMA 1.0 does not explicitly distinguish between process noise and observation noise; an interesting direction for further research would be to incorporate an explicit observation model.

Two ongoing challenges in Bayesian computation relevant to CheMA include inference of model parameters and  computation of marginal likelihoods for model selection. 
The first has been tackled from many directions, including approximate Bayesian computation \citep{Toni}, Gaussian processes approximations \citep{Dondelinger2}, MCMC \citep{Wilkinson}, and particle filtering \citep{Quach}. 
The second question is a comparatively under-developed area of statistical research, with candidate approaches including variational approximations \citep{Rue} and MCMC \citep{Vyshemirsky}.
In general the computational burden of CheMA will be higher than many methods (see Supplementary Information).
By way of demonstration, Bayesian inference and prediction for a system of 27 protein species required over 12 hours (serial) computational time. 
In contrast, linear or discrete models offer improved scalability to high-dimensional systems by permitting closed form expression of model selection criteria. 
Thus, CheMA can complement existing methodologies but is not at present applicable to truly high-dimensional problems with hundreds or thousands of nodes.


Finally we note the following caveats: (i) The automatic generation of kinetic equations  limits the extent to which in-depth knowledge about particular biochemical processes and dynamics may be incorporated. (ii) Our empirical results suggest that more complex interactions, including kinase inhibition, can be extremely difficult to identify from time series data. (iii) The form of kinetics used here will likely be sub-optimal when the assumptions of the Michaelis-Menten approximation are violated.
(iv) Larger training and test datasets may be needed to allow truly effective trajectory prediction and comprehensive assessment of performance.

\section*{Acknowledgement}

This work was supported by the US Department of Energy [DE-AC02-05CH11231]; US National Institute of Health, National Cancer Institute [U54 CA 112970, P50 CA 58207]; UK Engineering and Physical Sciences Research Council [EP/E501311/1]; and Netherlands Organisation for Scientific Research [Cancer Systems Biology Center].

\end{document}